# A large new family of filled skutterudites stabilized by electron count

Huixia Luo[1*], Jason W. Krizan[1], Lukas Muechler[1,4], Neel Haldolaarachchige[1], Tomasz Klimczuk[2], Weiwei Xie[1], Michael K. Fuccillo[1], Claudia Felser[4,5] and Robert J. Cava[1*]

[1]Department of Chemistry, Princeton University, Princeton, New Jersey 08544, USA

[2]Faculty of Applied Physics and Mathematics, Gdansk University of Technology, Narutowicza 11/12, 80-233 Gdansk, Poland

[4]Max-Planck-Institut für Chemische Physik Fester Stoffe, 01187, Dresden, Germany

[5]Johannes Gutenberg University, Institute of Inorganic Chemistry and Analytical Chemistry, Staudingerweg 9, D-55128 Mainz

* rcava@princeton.edu; huixial@princeton.edu

**Based on the interplay of theory and experiment, a large new family of filled group 9 (Co, Rh and Ir) skutterudites is designed and synthesized. The new materials fill the empty cages in the structures of the known binary $CoSb_3$, $RhSb_3$ and $IrSb_3$ skutterudites with alkaline, alkaline earth, and rare earth atoms to create compounds of the type $A_yB_4X_{12}$; A atoms fill the cages to a fraction y, B are the group 9 transition metals, and X is a mixture of electronegative main group elements chosen to achieve chemical stability by adjusting the electron counts to electron-precise values. Forty-three new compounds are reported, antimony-tin and phosphorous-silicon based, with 63 compositional variations presented. The new family of compounds is large and general. The results described here can be extended to the synthesis of hundreds of new group 9 filled skutterudites.**

The Zintl phases are a family of solid compounds typically made from electropositive metallic elements combined with non-metals or metalloids capable of forming polyanions[1]. The more electropositive elements donate their valence electrons to the more electronegative elements, and the electronegative elements form polyanions to accept the donated electrons. The bonding and architecture of the polyanions is understood in terms of the octet rule.[2-4] Due to the filled valence states of the polyanion array, and the empty valence states of the electropositive electron donors, semiconducting behavior is favored for the resulting compounds.

The binary skutterudites are one well known class of Zintl compounds, deriving from the archetypal mineral skutterudite, $CoAs_3$[4,5]. They have the general formula $BX_3$ where $B$ is a transition metal element such as Fe, Co, Rh or Ir, and $X$ is a pnictogen such as P, As, or Sb (see Figure 1a)[5-7]; they typically crystallize in cubic space group $Im\bar{3}$ (#204) with two $B_4X_{12}$ formula units and two large empty cages per unit cell. The B ions are in the 8c (1/4, 1/4, 1/4) site and the X ions are in the 24g (0, y, z) site with y ~0.15 and z ~ 0.35. Featured in the structure are distorted square $X_4$ Zintl polyanions that have a formal charge of 4-. Thus the simple binary compounds have a semiconducting electron count for $B$ elements from group 9; the $B$ ions, Co, Rh and Ir, formally 3+, have the electron configuration $nd^6$. In octahedral coordination and low spin, they thus have a filled $t_{2g}$ band and an empty $e_g$ band, with an energy gap between them. $CoAs_3$ for example can be understood as $Co^{3+}_4(As_4)^{4-}_3$[8-10].

The empty cages within the binary skutterudite framework can be filled by up to one ion ($A$) per $B_4X_{12}$ formula unit, leading to the "filled skutterudites" of formula $A_yB_4X_{12}$, where $A$ = alkali, alkaline-earth, rare-earth, actinide, or Tl (see Figure 1b)[11-13]. These are also in space group #204, with the A ions (in the ideal structures) in the 2a (0,0,0) site and the B and X ions as in the unfilled case. The known filled skutterudites are virtually all based on the group 8 metals Fe, Ru and Os. The *A* ion donates its charge to the $B_4X_{12}$ framework. By judicious choice of constituents, complete filling of the cages ($y = 1$), a Zintl (electron precise) electron count, and thus semiconducting behavior can be achieved. In addition to those with an electron precise formula some compounds such as $LaRu_4Sb_{12}$ (with one electron in deficit of the Zintl count) are also known. A very small number of filled skutterudites based on Co, Ni and Pt have been reported (see Figure 1c)[15-50]. Because they make very good thermoelectrics[14,22-26], host unusual metal-insulator transitions[27-30], and can be rare earth magnets, heavy-fermion compounds[31-34], non-Fermi liquids[35-39], itinerant ferromagnets[40-42], superconductors[31, 43-50], and have been proposed as topological insulators[51], filled skutterudites are a very important class of non-molecular solids. Because the electronic and magnetic properties of solids depend critically on the actual atoms making up the compounds in addition to the compound's electron count, expanding the filled skutterudites from primarily group 8 to group 9 metal based compounds, as we have done in this study, affords the opportunity to observe many new physical properties.

Here we report that an extremely large family of filled skutterudites based on the group 9 metals Co, Rh and Ir can be chemically stabilized if the electron count is stabilized by partial substitution on the X ion site to yield electron-precise formulas. These new filled skutterudites were designed by filling or partial filling the cages in the binary $CoSb_3$, $RhSb_3$ and $IrSb_3$ frameworks with alkali (Li, Na, K), alkaline earth metal (Ca, Sr, Ba), and rare earth atoms (La, Ce, Pr, Nd, Gd, Yb), by compensating for the extra positive charge of the filling ion by partial substitution of Sn on the Sb site or Si on the P site to yield electron-precise formulas, guided by the predictions of first principles electronic structure calculations. The pure antimonides and pure phosphides (with no Sb/Ns or P/Si mixing) are not stable, but we find that electron precise formulas are not strictly required for compound stability in some cases. In addition to their synthesis, crystal structures and electronic structures, we briefly survey the magnetic properties of selected members of this large new family of compounds.

**Results and Discussion**

The prototypical binary $BX_3$ skutterudites with B = Co, Rh, and Ir are non-magnetic semiconductors, with a Zintl electron count of 24 valence electrons per formula unit. (The 24 electron count rule per $BX_3$ unit for semiconducting behavior also holds for a recently proposed non-Zintl view of the formal electron configuration of the constituent elements in skutterudites[52].) Filled $AB_4X_{12}$ skutterudites that have B in a $d^6$ low spin configuration and a $p^6$ X atom configuration are expected to be at electron precise, semiconducting, non-magnetic compounds when they have a valence electron count of 96 (4 x 24) per formula unit. The power of the Zintl concept is that these simple electron counting rules enable the prediction of new thermodynamically stable compounds. A Cobalt-based filled skutterudite compound can be stabilized, for example, by using Ba as an $A^{2+}$ ion and compensating for the added electrons by removing 2 electrons from the X ion site, and thus $BaCo_4Sb_{10}Sn_2$ should be stable, and an electron precise, non-magnetic semiconductor. In the same fashion the Rh and Ir compounds can be obtained as can the filled skutterudite semiconductors for any A ion.

This simple electron count principle is confirmed by our ab-initio calculations for several examples in this family, which are shown in Figure 2a. The valence band consists of strongly hybridized metal-d and Pnictogen-p states and is separated by a gap from the conduction band showing the expected semiconducting character. Surprisingly, the electronic structure of the Rh variant calculated with the GGA is semimetallic, with a small valence band - conduction band overlap. Use of the MBJ functional in the calculation, which accounts for electronic correlations more effectively, on the other hand opens a band gap and predicts the Rh variant to be a semiconductor, although the predicted band gap is very small. This predicted anomalous behavior of the filled skutterudite variants based on Rh, which can be attributed to special characteristics of the Rh-Sb electronic interactions, would be of interest to verify through experiment.

The electron counting principle employed is similarly supported by the valence orbital region of the electronic DOS curves for the hypothetical compounds with formulas (simplified to allow for the calculations through the omission of Sn/Sb mixing) "La[Co/Rh/Ir]$_4$Sb$_{12}$" (Figure 2b) obtained by using the TB-LMTO-ASA calculations. There are two distinct regions: (i) the region 6eV below the Fermi level ($E_F$) shows mostly Sb 5s bands mixed with s electrons from the transition metals [Co/Rh/Ir] and small amounts of La valence orbitals; and (ii) the region 0-6 eV below $E_F$ arises mostly from Sb 5p bands mixed with d electrons from Co/Rh/Ir states. There is a clear gap below $E_F$ at 3 electrons per formula unit lower than the model compounds (which did not have Sn partially substituting for Sb). Similarly, analogous calculations for K and Ba filled skutterudites with no X site partial substitution show band gaps 1 and 2 electrons below $E_F$, respectively. This is as expected from the Zintl counting rule. In line with the Zintl-Klemm formalism, the La–Sb -COHP curves indicate maximum orbital interactions, i.e., bonding and covalency, in La[Co/Rh/Ir]$_4$Sb$_{12}$ and La[Co/Rh/Ir]$_4$Sb$_9$Sn$_3$. In contrast, strong antibonding contributions come from La-Co/Rh/Ir interactions at Fermi level in both

La[Co/Rh/Ir]$_4$Sb$_{12}$ and La[Co/Rh/Ir]$_4$Sb$_9$Sn$_3$. This indicates that the cations prefer not to be close to each other, which may be the reason why these and other filled skutterudites can show A site deficiencies under various synthetic conditions.

The compositions of the filled skutterudites was determined by standard solid state phase equilibria methods, which, using the characterization of reaction products by powder X-ray diffraction, especially for the heavy elements in compounds such as these, is quite sensitive to the presence of impurities. In addition, the crystal structures of all 63 variants reported here were determined by quantitative fits to the powder diffraction patterns. Figures 3 show the powder XRD patterns for the quantitative Rietveld structure refinements for 12 examples of compounds the Co-series, Rh-series, Ir-series and P-series filled skutterudites. The agreement between the observed and calculated patterns in all cases is excellent; a selection of the structural refinements are shown in Table 1 with the remainder given in the supplemental information in Table 1S. The cubic crystallographic cell parameters for all the compositions synthesized, determined by least squares fitting to 20 or more observed reflections in the powder XRD patterns using profile fits, are found in Table 2.

Figure1S shows the Rietveld refinements of the laboratory powder XRD data for the filled skutterudites made in this work. The ideal filled skutterudite structure is primarily found, with random Sb/Sn and P/Si mixing on the x ion site. For all the filled skutterudites based on P, and for the Sb-based variants based on Co, the *A* ions are found in the centers of the skutterudite cages (the 2a, (0,0,0) site), their ideal positons, as is frequently assumed for filled skutterudites in other chemical families. It is of interest, however, that for the compounds with the larger *B* ions, and consequently larger unit cells and X cages, we find that the *A* ions are sometime displaced from the centers of the cages, i.e. off the ideal 2a site. This is found for the

Ir compounds in particular. The qualitative evidence for this displacement is seen in comparing Figure 3a to Figures 3b and 3c; it can be seen that the intensities of the low angle diffraction peaks are suppressed by when the Co-skutterudites are filled, but that there is residual intensity seen in the Rh and Ir skutterudites. The three models evaluated to best model the observed diffraction peak intensities, rather than employing the ideal filled skutterudite structure, were: (a) refine the 2a site occupancy, (b) refine the 2a site atomic displacement parameter, and (c) move the A atom off the ideal 2a site toward the surrounding X cage. (a) is not appropriate because the samples are single phase at the stoichiometry determined by phase equilibria studies; the formulas were further confirmed by Energy dispersive X-ray spectroscopy (EDS). (b) provides an adequate fit, yielding excessively large atomic displacement parameters, implying that the A site ions are dynamically displaced from their ideal positions. The alternative (c) where the A atom is randomly displaced to partially occupy sites along a <111> type direction to minimize the A-X distances in the cage, seems like the most physically realistic scenario and was the model used here to describe the rare earth ion disorder. In a handful of cases, $BaIr_4Sb_{10}Sn_2$ and $La_{0.9}Ir_4Sb_{10.2}Sn_{1.8}$, for example realistic off-center A atom positions cannot sufficiently fully account for the observed intensities of the low angle peaks, and thus it may be that a small fraction of the A ions are substituted on the B atom sites. Although the existence of $5d$ Os and Pt rare earth filled skutterudites been reported in the past[9,10,53,54], few quantitative structure determinations have been reported. A model like (b), with anomalously large atomic displacement parameters for the lanthanides, has been used to describe the Ln site disorder for Os-based pyrochlores[55]. The very low thermal conductivity of the filled skutterudites based on group 8 elements has been attributed to the anharmonic thermal vibration of the *A* site ion "rattling" in the skutterudite cage[14, 56], implying a tendency toward off-center positioning of the A site ion in the cages in this family, in agreement with our structural refinements and those where the A site ion disorder is modelled by large thermal parameters. It is not yet known whether the A site disordered position model we observe here is a general feature of all larger filled skutterudites. Whether the disorder is static or dynamic or a combination of both, and whether there is any complex defect chemistry present in some of these compounds, would be of interest to study further by other structural characterization methods.

The important structural features of the filled skutterudites studied here are presented in Figure 4. This figure shows the geometry of the A site and group 9 site polyhedra and the manner in which they share faces (leading to the strong antibonding interactions described above), the displacements of the A site ions from the cage centers, and the $X_4$ squares. The lower part of the figure compares the $X$-$X$ bond lengths within the slightly distorted $X_4$ antimonide and phosphide squares and the $B$-$X$ bond lengths (all 6 are equivalent) for several representative members of the group 9 filled skutterudites. Neither the $X_4$ "squares" nor the $BX_6$ octahedra are constrained by symmetry to be perfect, and they are not, though they are all close and the distortions are relatively small. Because the same electron count is nearly perfectly maintained for all the compounds studied quantitatively here, large bond-length differences in these critical structural components are not expected to occur.

Filled skutterudites based on rare earth metals and group 8 transition elements show diverse and interesting rare earth magnetism, and they have been widely studied; see e.g [32,33,35]. In order to generally survey the magnetic properties for our new filled group 9 based skutterudites, the magnetizations of all the rare earth based filled skutterudites were measured from 2 to 200 K with an applied magnetic field of

$\mu_0 H = 1$ Tesla. The compounds are strongly paramagnetic as expected due to the rare earth magnetism. We consider the analysis of several set of compounds in more detail as examples. Figure 5a, b and c show in the main panel the inverse magnetic susceptibilities ($\chi$ is defined as $M/\mu_0 H$) for the $Ce_{0.9}[Co/Rh/Ir]_4Sb_{10.2}Sn_{1.8}$ $Yb_{0.9}[Co/Rh/Ir]_4Sb_{10.2}Sn_{1.8}$ and $Pr_{0.9}[Co/Rh/Ir]_4Sb_{10.2}Sn_{1.8}$ samples. The temperature dependences of the inverse magnetic susceptibilities, corrected for $\chi_0$, are linear at high temperatures for all three compounds, indicating Curie Weiss behavior. Fits to the susceptibilities were performed in the temperature range of 100-200K, according to $\chi-\chi_0 = C/(T-\theta_{cw})$, where C is the Curie constant, $\theta_{cw}$ is the paramagnetic Curie temperature, and $\chi_0$ is the temperature independent contribution to the susceptibility. From these fits, the effective magnetic moment ($P_{eff}$) per Ln (Ln = La, Ce, Pr, Nd, Gd, and Yb) ion was obtained by using $P_{eff} = (8C)^{1/2}$. The thus derived basic magnetic characteristics (i.e. $\theta_{cw}$ and $P_{eff}$) for all our new compounds containing magnetic rare earths are summarized in Table 2. Fitting the magnetic susceptibility using the Curie Weiss law in the range 100 to 200 K, we obtain the effective moments for $Pr_{0.9}Co_4Sb_{10.2}Sn_{1.8}$, $Pr_{0.9}Rh_4Sb_{10.2}Sn_{1.8}$, and $Pr_{0.9}Ir_4Sb_{10.2}Sn_{1.8}$ of $P_{eff}$ = 3.43, 3.66 and 3.66 $\mu_B$/Pr, respectively, close to the effective moment value expected for the free $Pr^{3+}$ free ion ($P_{eff}$ = 3.58 $\mu_B$/Pr). In addition, the magnetic susceptibility data show a broad peak at around 3.5 K for $Pr_{0.9}Rh_4Sb_{10.2}Sn_{1.8}$ (Figure 5d), which implies the onset of antiferromagnetic ordering. In order to better estimate the Neel temperature, we follow standard procedure [57-59] and plot $d(\chi T)/dT$ (inset, Figure 5d). The maximum of $d(\chi T)/dT$ is observed at 3.5 K, which can be taken as $T_N$. Similar fits were performed for all the rare earth compounds synthesized.

The effective moments for the $Ce_{0.9}[Co/Rh/Ir]_4Sb_{10.2}Sn_{1.8}$ samples are in the range $P_{eff}$ = 2.1-2.5 $\mu_B$/Ce, which are close to the expected free-ion Hund's rule value of

2.54 $\mu_B$/Ce$^{3+}$. These values are similar to the effective moment ($P_{eff}$ = 2.1-2.26 $\mu_B$/Ce) reported for the group 8 filled skutterudites CeRu$_4$Sb$_{12}$[36], but are different from those reported for CeFe$_4$Sb$_{12}$ ($P_{eff}$ = 3.5-3.8 $\mu_B$/Ce, which is larger than the value expected for a Ce$^{3+}$ [60,61]). For the Yb-filled phases, $P_{eff}$ = 1.56, 2.76 and 2.71 $\mu_B$/Yb for Yb$_{0.9}$Co$_4$Sb$_{10.2}$Sn$_{1.8}$, Yb$_{0.9}$Rh$_4$Sb$_{10.2}$Sn$_{1.8}$, and Yb$_{0.9}$Ir$_4$Sb$_{10.2}$Sn$_{1.8}$, respectively. For all three samples, the effective moments are intermediate between those for Yb$^{3+}$ ($P_{eff}$ = 4.54 $\mu_B$/Yb for the free ion) and nonmagnetic Yb$^{2+}$ ($P_{eff}$ = 0). Depressed values of effective moment for Yb have also been seen in YbFe$_4$Sb$_{12}$ ($P_{eff}$ = 3.09 $\mu_B$/Yb)[62] and Yb compounds in general show complex temperature dependent susceptibilities due to the energetic proximity of the *f* level excited states to the ground state and hybridization of the Yb *f* orbitals with other electronic states.

The new filled group 9 skutterudites that do not contain rare earths display temperature independent susceptibilities, with the exception of weak "Curie tails" at low temperatures due to the presence of impurity spins. Almost all of the intrinsic susceptibilities are in fact diamagnetic for these materials, indicating that they are dominated by core diamagnetism, as expected from the electron-precise formulas. The magnetic susceptibilities at 150 K for the non-magnetic samples, also presented in Table 2, are between -0.001 and -0.01 emu per mol formula unit. This indicates that the compounds do not display local moment behavior, but rather band behavior, even for the Co variants. More detailed study of the magnetic properties of the group 9 filled skutterudites will be of future interest.

**Calculation, Synthesis and Experimental**

The electronic band structure calculations were performed in the framework of Density Functional Theory (DFT) using the WIEN2k[53] code with a full-potential linearized augmented plane-wave and local orbitals [FP-LAPW + lo] basis together with the Perdew Burke Ernzerhof (PBE) parameterization[54] of the generalized gradient approximation (GGA) as the exchange-correlation functional. In one case (see text) the MBJ[63] functional was also used. The plane wave cut-off parameter $RK_{MAX}$ was set to 7 and the Brillouin zone was sampled by 500 $k$ points. To simulate the substitution of Sb by Sn, the virtual crystal approximation (VCA)[64,65] was employed. The density of states (DOS) and crystal orbital Hamilton population (COHP)[66] were generated by tight-binding linear-muffin-tin-orbital atomic-sphere-approximation (TB-LMTO-ASA) calculations using the Stuttgart code[67]. Exchange and correlation were treated by the local density approximation (LDA)[68]. In the ASA method, space is filled with overlapping Wigner-Seitz (WS) spheres. The empty spheres were necessary in the calculation, and the WS sphere overlap was limited to no larger than 16%. The basis set for the calculations included La (6s, 6p, 5d, 4f), Co (4s, 4p, 3d)/Rh (5s, 5p, 4d)/Ir (6s, 6p, 5d) and Sb (5s, 5p) wavefunctions. The convergence criterion was set to $10^{-4}$ eV. A mesh of 6 × 6 × 6 k-points in the irreducible wedge of the first Brillouin zone was used to obtain all values. Experimental lattice constants were used and the free internal parameters were optimized by minimizing the forces.

Single phase polycrystalline samples were synthesized from starting compositions $A$[Co/Rh/Ir]$_4$Sb$_{11}$Sn for $A^{1+}$ ions ($A$ = Li, Na, K), $A$[Co/Rh/Ir]$_4$Sb$_{10}$Sn$_2$ for $A^{2+}$ ions (A = Ca, Sr, Ba) and $A_{0.9}$[Co/Rh/Ir]$_4$Sb$_{10.2}$Sn$_{1.8}$ for $A^{3+}$ ions (A = Ln = La, Ce, Pr, Gd, and Yb). (Due to the large number of compounds described here, the formulas for the groups are abbreviated through the use of the square brackets for the group 9 element - in each case, pure Co, pure Rh, and pure Ir variants were made.) The single phase compositions reported here were determined in all cases through performing a series of systematic syntheses in extensive phase equilibria studies. The use of phase equilibria studies facilitated by powder X-ray diffraction is a powerful, common method for determining the compositions of solids. The single phase compositions reported here are those that yielded filled skutterudite compounds with $y$ closest to 1 in $A_yB_4X_{12}$. Samples were prepared by a solid state reaction method. The alkali- and alkaline-earth-based groups are single phase for starting compositions at the ideal Zintl electron count, but for the lanthanide-antimony based compounds, a more complex formula, with electrons in excess of the electron-precise count, was needed to make single phase materials under our conditions. For $A^{1+}$[Co/Rh/Ir]$_4$Sb$_{11}$Sn and $A^{2+}$[Co/Rh/Ir]$_4$Sb$_{10}$Sn$_2$, stoichiometric quantities of high-purity elemental Li/Na/K (99.999 %), Ca/Sr/Ba (99.9 %), Co (99.8 %), Rh (99.99 %) or Ir (99.99 %), Sb (99.999 %) and Sn (99.8 %) were mixed, pressed into a pellet, put into an alumina crucible, and then sealed in clean evacuated quartz ampoules. The ampoules were slowly heated up to 600 °C and held for 10 h, then slowly heated up to 700 °C and held for 10 h. Then they were removed from the furnace, thoroughly ground into powder, repressed into pellets, and reheated at 700 °C for 10 h. This process was repeated two to four times until single phase material was obtained. Critically, especially for the case of alkali filling of the skutterudite cavities, where a naive view might speculate that volatility or tube reaction might disrupt the synthesis, the internal surfaces of the quartz ampoules were completely clean after the syntheses, showing no sign of attack or mass loss during the synthetic procedure. All the mixing and grinding processes were performed in a glove box ($P_{O2}$ and $P_{H2O}$ < 1 ppm). For comparison purposes, the binary skutterudites CoSb$_3$, RhSb$_3$ and IrSb$_3$ were prepared by the same method. For Ln$_{0.9}$[Co/Rh/Ir]$_4$Sb$_{10.2}$Sn$_{1.8}$ (Ln = La, Ce, Pr, Gd, and Yb), precursor LnSn binaries or Ln[Co/Rh/Ir]Sn ternaries were first made by arc-melting the elements. The as-prepared binaries and ternaries were then ground into powder and mixed with the appropriate stoichiometric quantities of high-purity elemental Co (99.8 %), Rh (99.99 %), Ir (99.99 %), Sb (99.999 %) and Sn (99.8 %) powder, and heated as described above. Compounds based on phosphorous as the most electronegative element were also made, but for the rare earth family only; Si or Ge were partially substituted for P to yield the Zintl electron count. For $Ln[Co/Rh/Ir]_4P_{12-x}[Ge/Si]_x$, (Ln = La, Ce, Pr and Nd), the precursor binaries or $Ln[Co/Rh/Ir]_4[Ge/Si]_x$ ternaries were first made by arc-melting the elements. Weight loss in this process was less than 1%. The as-prepared binaries or ternaries were then ground into powder and mixed with the appropriate stoichiometric quantities of elemental red P (99.9 %), pressed into pellets and then sealed in clean evacuated quartz ampoules. These ampoules were slowly heated to 400 °C, held for 10 h, and then slowly heated to 700 °C and held for 10 h; they were then slowly heated to 900 °C and held overnight. Finally, they were thoroughly ground into a powder, repressed into pellets, and reheated at 900 °C and held there overnight.

The single phase compositions determined in the phase equilibria studies were specified by powder X-ray diffraction (XRD) using a Bruker D8 Focus diffractometer with Cu Kα radiation and a graphite diffracted beam monochromator. The initial structural model for the structures that were quantitatively refined from the powder diffraction data was taken from that of $LaRu_4Sb_{12}$[69]. The FullProf software suite was used for the Rietveld refinements. Peak shapes were modeled with the Thompson-Cox-Hastings pseudo-Voight profile convoluted with axial divergence asymmetry. The background was modeled with a Chebychev Polynomial. The structures were refined in space group $Im\bar{3}$ with Co, Rh or Ir in the 8c sites and Sb/Sn in the 24g sites. The position of the A site ion was dependent on the specific compound, as described.

In the final structural models, the structural parameters refined were the two positional coordinates (x and y) of the X atoms in position 24g, and when necessary, the position of the atom on the *A* atom site along the <111> direction and a small percent of site mixing of A from the 2a onto the 8c site; all other structural parameters are fixed by symmetry. The formulas observed in the refinements were all consistent with the compositions determined from the phase equilibria studies. Similarly, the compositions of representative compounds tested by EDX were consistent with those compositions. Measurements of the temperature dependence of the electrical resistivity and magnetization were performed in a Quantum Design Physical Property Measurement System (PPMS) from 2 to 300 K.

**Conclusion**

In conclusion, we have shown the existence of a very large new family of filled skutterudites based on the group 9 skutterudites $CoSb_3$, $RhSb_3$ and $IrSb_3$, stabilized by X site substitution to yield compounds with electron-precise formulas. The compounds were found through a combination of experimental studies and theoretical first principle calculations. In this work, 63 new filled skutterudite compositions based on the group 9 metals are reported, but by simple extension, for example to the full 14 member rare earth family, and to phosphides and arsenides in addition to the antimonides, we expect that several hundred new group 9 based filled skutterudite compounds can be found and characterized based on the concept presented here. This greatly expands the family of known filled skutterudites, an important solid structural family. The physical properties briefly surveyed here suggest that the rare earth-based materials will likely display interesting magnetic behavior deserving of more detailed study. The group 9 based filled skutterudites found here should display properties in some cases that are quite different from the commonly studied group 8 materials. The general approach employed – to search for stable solid compounds of interesting transition element ions, even complex ones, by aiming at electron-precise (i.e. Zintl) formulas, is likely to be a fruitful approach for expanding other families of solid compounds in the future.

**Acknowledgment**

This work was supported by the AFOSR MURIs in thermoelectric and superconducting materials, grants FA9550-10-1-0553 and FA9550-09-1-0953. The authors also greatly acknowledge Brendan F. Phelan for stimulating discussions.

**Author Contributions**

R.J.C, H.X.L, L.M., N.H and T.K. conceived and designed the experiments and H.X.L, T.K. and N.H performed the synthetic experiments. H.X.L. and R.J.C. analyzed and interpreted data. T.K. performed the magnetic characterization. J.K. performed and analyzed the XRD refinement data. C.F., L.M and W.W.X performed and analyzed the calculation data. M.K.F supported the Seebeck measurement and designed the TOC. H.X.L and R.J.C wrote the paper. All authors approved the content of the manuscript.

**Competing financial interests:** The authors declare no competing financial interest.

**Figures legends**

Figure 1. (a) Flowchart for designing new skutterudites using the Zintl concept. (b) Structures of $BX_3$ and $AB_4X_{12}$ showing the covalently bonding regions (red atoms) separate from the ionic regions (blue and green atoms). The atoms in the two regions can be partially substituted for tuning of the electron concentration to change the electronic structure. (c) Periodic table of the elements highlighted to show the new stuffed skutterudites found in this work and those reported previously.

Figure 2. Left column: calculated electronic structures of the filled skutterudites $LaCo_4Sb_9Sn_3$, $LaRh_4Sb_9Sn_3$, $LaIr_4Sn_9Sn_3$; Right column: results of TB-LMTO-ASA (LDA) [58,59] electronic structure calculations on $La[Co/Rh/Ir]_4Sb_{12}$ model compounds with La on 2a sites, Co/Rh/Ir on 8c sites and Sb/Sn on 24g sites.: (left) total DOS curve (right) -COHP curves [57] among all nearest (red) and second nearest (blue) neighbor interatomic contacts with La atom (+ is bonding and − is antibonding), the dashed black lines indicate the Fermi level for $La[Co/Rh/Ir]_4Sb_{12}$ and the dashed red lines indicate the Fermi level for $La[Co/Rh/Ir]_4Sb_9Sn_3$.

Figure 3. The Rietveld refinements of the laboratory powder XRD data for (a) $CoSb_3$, $K_{0.8}Co_4Sb_{11}Sn$, $BaCo_4Sb_{10}Sn_2$ and $LaCo_4Sb_{10.2}Sn_{1.8}$ (b) $RhSb_3$, $K_{0.8}Rh_4Sb_{11}Sn$, $BaRh_4Sb_{10}Sn_2$ and $La_{0.9}Rh_4Sb_{10.2}Sn_{1.8}$ (c) $IrSb_3$, $K_{0.8}Ir_4Sb_{11}Sn$, $BaIr_4Sb_{10}Sn_2$ and $LaIr_4Sb_{10.2}Sn_{1.8}$ (c) $LaCo_4P_9Ge_3$, $LaRh_4P_8Ge_4$, $LaIr_4P_9Ge_3$ and $LaCo_4P_9Si_3$. Red points observed data, black curves calculated pattern, green tics expected peak positions, lower blue curve, difference between observed and calculated diffraction pattern. Refined structural parameters found in Table 1.

Figure 4. Upper figure; Detail of the $AX_{12}$ polyhedron, the $BX_6$ octahedron, the A site displacement, the $X_4$ squares, and the manner in which they are related in the $La_{0.9}Ir_4Sb_{10.2}Sn_{1.8}$ filled skutterudite. Lower figure: Comparisons of the $X_4$ square and $BX_6$ octahedra geometries for selected filled group 9-based skutterudite compounds, from the current structure refinements.

Figure 5. Magnetic properties and zero-field cooled temperature-dependent dc inverse magnetic susceptibilities of Pr-filled group 9 based skutterudites: (a) $Ce_{0.9}[Co/Rh/Ir]_4Sb_{10.2}Sn_{1.8}$; (b) $Yb_{0.9}[Co/Rh/Ir]_4Sb_{10.2}Sn_{1.8}$; (c) $Pr_{0.9}[Co/Rh/Ir]_4Sb_{10.2}Sn_{1.8}$; (d) Enlarged low temperature region of the magnetic susceptibility of $Pr_{0.9}Rh_4Sb_{10.2}Sn_{1.8}$. Inset (upper): enlarged low temperature region of the inverse susceptibility. Inset (lower): $d(\chi T)/dT$.

Table 1. Structural parameters for refined crystal structures of selected filled skutterudites (and comparison to binary skutterudites).

Table 2. Cubic cell parameters and selected physical properties of group 9-based filled skutterudites.

Table 1S. Structural parameters for refined crystal structures of filled skutterudites (and comparison to binary skutterudites).

Figure1S. The Rietveld refinements of the laboratory powder XRD data for filled skutterudites (and comparison to binary skutterudites).

Figure 1. (a) Flowchart for designing new skutterudites using the Zintl concept. (b) Structures of $BX_3$ and $AB_4X_{12}$ showing the covalently bonding regions (red atoms) separate from the ionic regions (blue and green atoms). The atoms in the two regions can be partially substituted for tuning of the electron concentration to change the electronic structure. (c)Periodic table of the elements highlighted to show the new stuffed skutterudites found in this work and those reported previously.

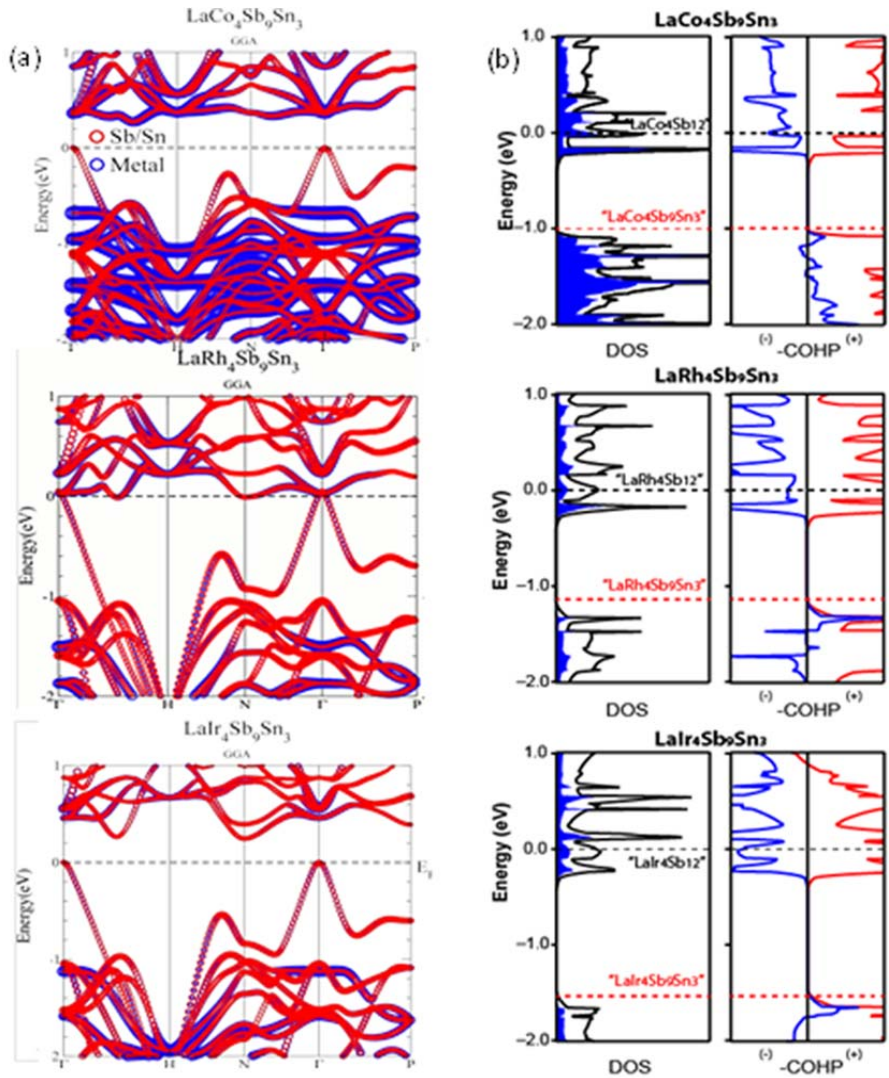

Figure 2. Left column: calculated electronic structures of the filled skutterudites LaCo$_4$Sb$_9$Sn$_3$, LaRh$_4$Sb$_9$Sn$_3$, LaIr$_4$Sn$_9$Sn$_3$; Right column: results of TB-LMTO-ASA (LDA) [58,59] electronic structure calculations on La[Co/Rh/Ir]$_4$Sb$_{12}$ model compounds with La on 2a sites, Co/Rh/Ir on 8c sites and Sb/Sn on 24g sites.: (left) total DOS curve (right) -COHP curves [57] among all nearest (red) and second nearest (blue) neighbor interatomic contacts with La atom (+ is bonding and − is antibonding), the dashed black lines indicate the Fermi level for La[Co/Rh/Ir]$_4$Sb$_{12}$ and the dashed red lines indicate the Fermi level for La[Co/Rh/Ir]$_4$Sb$_9$Sn$_3$.

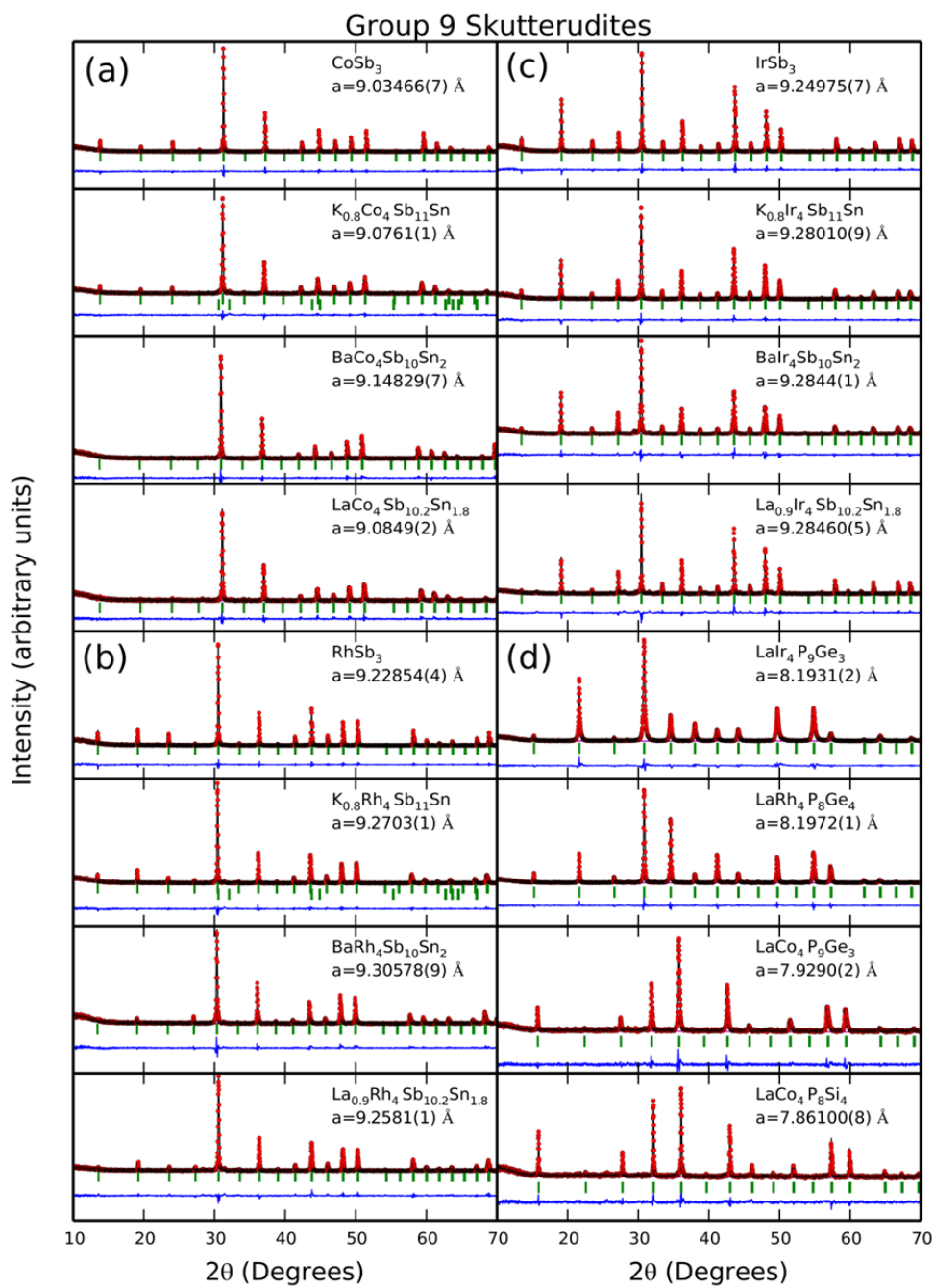

Figure 3. The Rietveld refinements of the laboratory powder XRD data for (a) $CoSb_3$, $K_{0.8}Co_4Sb_{11}Sn$, $BaCo_4Sb_{10}Sn_2$ and $LaCo_4Sb_{10.2}Sn_{1.8}$ (b) $RhSb_3$, $K_{0.8}Rh_4Sb_{11}Sn$, $BaRh_4Sb_{10}Sn_2$ and $La_{0.9}Rh_4Sb_{10.2}Sn_{1.8}$ (c) $IrSb_3$, $K_{0.8}Ir_4Sb_{11}Sn$, $BaIr_4Sb_{10}Sn_2$ and $LaIr_4Sb_{10.2}Sn_{1.8}$ (c) $LaCo_4P_9Ge_3$, $LaRh_4P_8Ge_4$, $LaIr_4P_9Ge_3$ and $LaCo_4P_9Si_3$. Red points observed data, black curves calculated pattern, green tics expected peak positions, lower blue curve, difference between observed and calculated diffraction pattern. Refined structural parameters found in Table 1. All patterns were collected with Cu-kα at 300 K.

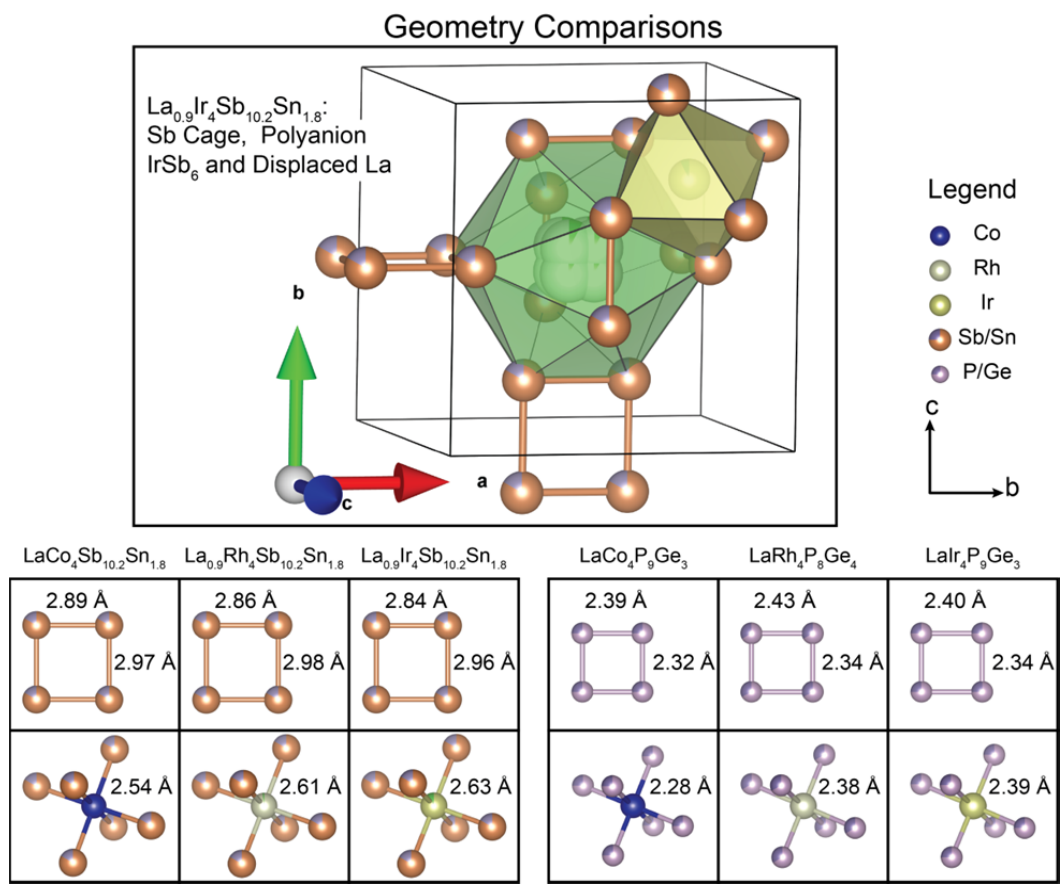

Figure 4. Upper figure; Detail of the $AX_{12}$ polyhedron, the $BX_6$ octahedron, the A site displacement, the $X_4$ squares, and the manner in which they are related in the $La_{0.9}Ir_4Sb_{10.2}Sn_{1.8}$ filled skutterudite. Lower figure: Comparisons of the $X_4$ square and $BX_6$ octahedra geometries for selected filled group 9-based skutterudite compounds, from the current structure refinements.

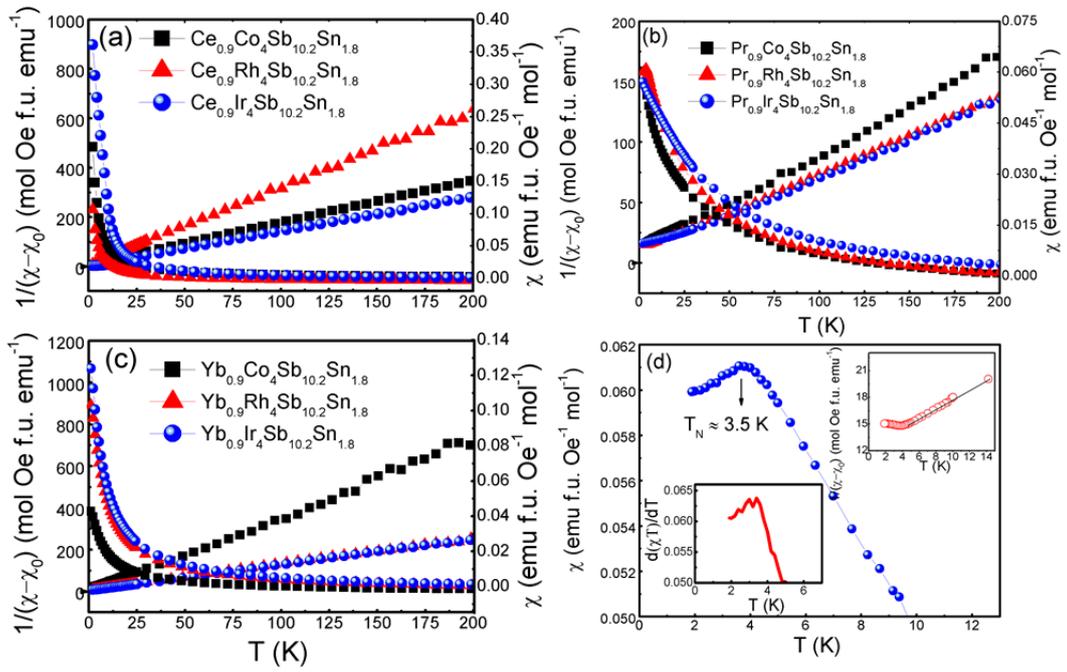

Figure 5. Magnetic properties and zero-field cooled temperature-dependent dc inverse magnetic susceptibilities of Pr-filled group 9 based skutterudites: (a) $Ce_{0.9}[Co/Rh/Ir]_4Sb_{10.2}Sn_{1.8}$; (b) $Pr_{0.9}[Co/Rh/Ir]_4Sb_{10.2}Sn_{1.8}$; (c) $Yb_{0.9}[Co/Rh/Ir]_4Sb_{10.2}Sn_{1.8}$; (d) Enlarged low temperature region of the magnetic susceptibility of $Pr_{0.9}Rh_4Sb_{10.2}Sn_{1.8}$. Inset (upper): enlarged low temperature region of the inverse susceptibility. Inset (lower): $d(\chi T)/dT$.

# Supplementary Information

## A large new family of filled skutterudites stabilized by electron count

Huixia Luo[1*], Jason W. Krizan[1], Lukas Muechler[1,4], Neel Haldolaarachchige[1], Tomasz Klimczuk[2], Weiwei Xie[1], Michael K. Fuccillo[1], Claudia Felser[4] and Robert J. Cava[1*]

[1]Department of Chemistry, Princeton University, Princeton, New Jersey 08544, USA

[2]Faculty of Applied Physics and Mathematics, Gdansk University of Technology, Narutowicza 11/12, 80-233 Gdansk, Poland

[4]Max-Planck-Institut für Chemische Physik Fester Stoffe, 01187, Dresden, Germany

\* rcava@princeton.edu;huixial@princeton.edu

**Table 1S**. Structural parameters for refined crystal structures of filled skutterudites (and comparison to binary skutterudites).

| Compound | $a_0$ (Å) | B-Site(y) | B-Site(z) | A-Site coordinate in (x,x,x) | $\chi^2$ |
|---|---|---|---|---|---|
| $CoSb_3$ | 9.03466(7) | 0.1577(3) | 0.3344(3) | N/A | 1.24 |
| $Li_{0.8}Co_4Sb_{11}Sn$ | 9.05908(8) | 0.1577(6) | 0.3339(6) | 0.80 | 1.51 |
| $Na_{0.8}Co_4Sb_{11}Sn$ | 9.06595(8) | 0.1580(4) | 0.3349(4) | 0.80 | 1.59 |
| $K_{0.8}Co_4Sb_{11}Sn$ | 9.0761(1) | 0.1590(4) | 0.3368(4) | 0.80 | 1.31 |
| $CaCo_4Sb_{11}Sn$ | 9.0634(1) | 0.1585(9) | 0.3356(9) | 1.00 | 1.52 |
| $BaCo_4Sb_{10}Sn_2$ | 9.14829(7) | 0.1626(6) | 0.3417(5) | 1.00 | 1.25 |
| $LaCo_4P_9Si_3$ | 7.86100(8) | 0.1505(9) | 0.3525(8) | 1.00 | 1.38 |
| $LaCo_4P_9Ge_3$ | 7.9290(2) | 0.1507(8) | 0.3538(7) | 1.00 | 1.5 |
| $LaCo_4Sb_9Sn_3$ | 9.1020(1) | 0.161(1) | 0.3373(10) | 1.00 | 1.59 |
| $LaCo_4Sb_{10}Sn_2$ | 9.0907(1) | 0.1605(8) | 0.3366(7) | 1.00 | 1.49 |
| $LaCo_4Sb_{10.1}Sn_{1.9}$ | 9.0811(2) | 0.1606(8) | 0.3372(7) | 1.00 | 1.59 |
| $LaCo_4Sb_{10.2}Sn_{1.8}$ | 9.0849(2) | 0.1589(6) | 0.3363(6) | 1.00 | 1.34 |
| $LaCo_4Sb_{10.3}Sn_{1.7}$ | 9.0843(1) | 0.1595(7) | 0.3365(6) | 1.00 | 1.55 |
| $CeCo_4P_9Ge_3$ | 7.9307(1) | 0.1521(10) | 0.3526(9) | 1.00 | 1.44 |
| $Ce_{0.9}Co_4Sb_{10.2}Sn_{1.8}$ | 9.0740(1) | 0.159(1) | 0.337(1) | 0.90 | 1.87 |
| $PrCo_4P_9Ge_3$ | 7.9171(1) | 0.151(2) | 0.351(1) | 1.00 | 1.96 |
| $Pr_{0.9}Co_4Sb_{10.2}Sn_{1.8}$ | 9.07274(10) | 0.1595(8) | 0.3356(8) | 1.00 | 1.54 |
| $NdCo_4P_8Ge_4$ | 7.9087(2) | 0.150(1) | 0.352(1) | 1.00 | 1.61 |
| $Yb_{0.9}Co_4Sb_{10.2}Sn_{1.8}$ | 9.0813(1) | 0.1593(8) | 0.3360(8) | 0.90 | 1.44 |
| $RhSb_3$ | 9.22854(4) | 0.1538(3) | 0.3395(3) | N/A | 1.36 |
| $Li_{0.8}Rh_4Sb_{11}Sn$ | 9.24843(5) | 0.1534(5) | 0.3388(5) | 0.80 | 1.75 |
| $Na_{0.8}Rh_4Sb_{11}Sn$ | 9.25748(6) | 0.1540(5) | 0.3399(5) | 0.80 | 1.4 |
| $K_{0.8}Rh_4Sb_{11}Sn$ | 9.2703(1) | 0.1546(4) | 0.3419(4) | 0.80 | 1.42 |
| $KRh_4Sb_{10}Si_2$ | 9.2484(1) | 0.1549(8) | 0.3409(8) | 1.00 | 1.93 |
| $KRh_4Sb_{10}Ge_2$ | 9.21706(6) | 0.1537(4) | 0.3421(4) | 1.00 | 1.42 |
| $KRh_4Sb_{10}Sn_2$ | 9.28241(8) | 0.1553(5) | 0.3418(5) | 1.00 | 1.41 |
| $CaRh_4Sb_{10.2}Sn_{1.8}$ | 9.27135(9) | 0.1533(9) | 0.3403(9) | 1.00 | 1.89 |
| $CaRh_4Sb_{11}Sn$ | 9.24925(6) | 0.1530(6) | 0.3396(6) | 1.00 | 1.59 |
| $SrRh_4Sb_{11}Sn$ | 9.2365(1) | 0.154(1) | 0.341(1) | 0.65 | 2.52 |
| $BaRh_4Sb_9Ge_3$ | 9.1970(1) | 0.1551(10) | 0.3428(10) | 0.72 | 2.2 |
| $BaRh_4Sb_{10}Sn_2$ | 9.30578(9) | 0.1569(6) | 0.3442(6) | 0.91 | 1.56 |
| $LaRh_4P_8Ge_4$ | 8.1972(1) | 0.1485(7) | 0.3572(6) | 1.00 | 1.97 |
| $LaRh_4Sb_8Sn_4$ | 9.27562(9) | 0.1587(8) | 0.3424(7) | 1.00 | 1.67 |
| $LaRh_4Sb_9Sn_3$ | 9.27196(7) | 0.1576(8) | 0.3424(8) | 1.00 | 1.5 |
| $La_{0.9}Rh_4Sb_{10.2}Sn_{1.8}$ | 9.2581(1) | 0.1542(5) | 0.3389(5) | 0.84 | 1.56 |
| LaRh$_4$Sb$_{11}$Sn | 9.2338(1) | 0.1540(7) | 0.3403(7) | 0.61 | 1.85 |
| CeRh$_4$P$_8$Ge$_4$ | 8.1918(1) | 0.148(1) | 0.356(1) | 1.00 | 2.55 |
| CeRh$_4$Sb$_8$Sn$_4$ | 9.26817(9) | 0.158(1) | 0.342(1) | 1.00 | 1.85 |
| CeRh$_4$Sb$_9$Sn$_3$ | 9.26825(8) | 0.1572(9) | 0.3419(9) | 1.00 | 1.56 |
| Ce$_{0.9}$Rh$_4$Sb$_{10.2}$Sn$_{1.8}$ | 9.26467(7) | 0.1553(7) | 0.3403(7) | 0.67 | 1.53 |
| PrRh$_4$P$_8$Ge$_4$ | 8.1709(2) | 0.148(1) | 0.3572(10) | 1.00 | 2.23 |
| PrRh$_4$Sb$_8$Sn$_4$ | 9.26743(8) | 0.157(1) | 0.3409(10) | 1.00 | 1.81 |
| PrRh$_4$Sb$_9$Sn$_3$ | 9.26684(8) | 0.1564(8) | 0.3404(8) | 1.00 | 1.62 |
| Pr$_{0.9}$Rh$_4$Sb$_{10.2}$Sn$_{1.8}$ | 9.25417(6) | 0.1545(6) | 0.3402(7) | 0.62 | 1.66 |
| NdRh$_4$P$_8$Ge$_4$ | 8.18336(7) | 0.1471(7) | 0.3564(7) | 1.00 | 2.06 |
| Gd$_{0.9}$Rh$_4$Sb$_{10.2}$Sn$_{1.8}$ | 9.27048(9) | 0.1541(10) | 0.340(1) | 0.65 | 1.91 |
| Yb$_{0.9}$Rh$_4$Sb$_{10.2}$Sn$_{1.8}$ | 9.26477(8) | 0.1548(10) | 0.3402(10) | 0.62 | 1.81 |
| IrSb$_3$ | 9.24975(7) | 0.1524(3) | 0.3396(3) | N/A | 1.28 |
| Li$_{0.8}$Ir$_4$Sb$_{11}$Sn | 9.26723(4) | 0.1528(4) | 0.3395(4) | 0.80 | 1.78 |
| Na$_{0.8}$Ir$_4$Sb$_{11}$Sn | 9.27078(5) | 0.1523(5) | 0.3410(5) | 0.80 | 1.53 |
| K$_{0.8}$Ir$_4$Sb$_{11}$Sn | 9.28010(9) | 0.1536(3) | 0.3414(3) | 0.80 | 1.28 |
| CaIr$_4$Sb$_{10.2}$Sn$_{1.8}$ | 9.28706(5) | 0.1529(5) | 0.3402(5) | 1.00 | 1.64 |
| CaIr$_4$Sb$_{11}$Sn | 9.26936(5) | 0.1524(4) | 0.3397(4) | 1.00 | 1.48 |
| SrIr$_4$Sb$_{11}$Sn | 9.25237(7) | 0.1524(8) | 0.3400(9) | 0.75 | 1.84 |
| BaIr$_4$Sb$_{10}$Sn$_2$ | 9.2844(1) | 0.1542(5) | 0.3420(5) | 0.49 | 1.55 |
| LaIr$_4$P$_9$Ge$_3$ | 8.1931(2) | 0.147(2) | 0.357(2) | 1.00 | 2.69 |
| LaIr$_4$Sb$_8$Sn$_4$ | 9.28871(8) | 0.1550(8) | 0.3409(8) | 0.85 | 1.72 |
| La$_{0.9}$Ir$_4$Sb$_{10.2}$Sn$_{1.8}$ | 9.28460(5) | 0.1530(7) | 0.3407(7) | 0.60 | 1.86 |
| CeIr$_4$P$_9$Ge$_3$ | 8.1860(1) | 0.146(2) | 0.357(2) | 1.00 | 3.17 |
| Ce$_{0.9}$Ir$_4$Sb$_{10.2}$Sn$_{1.8}$ | 9.26946(9) | 0.1527(8) | 0.3406(8) | 0.61 | 1.94 |
| PrIr$_4$P$_9$Ge$_3$ | 8.19980(7) | 0.144(1) | 0.358(1) | 1.00 | 2.23 |
| PrIr$_4$Sb$_8$Sn$_4$ | 9.29333(7) | 0.1560(8) | 0.3419(8) | 1.00 | 1.62 |
| Pr$_{0.9}$Ir$_4$Sb$_{10.2}$Sn$_{1.8}$ | 9.27728(5) | 0.1531(5) | 0.3406(6) | 0.61 | 1.5 |
| NdIr$_4$P$_9$Ge$_3$ | 8.19613(9) | 0.143(1) | 0.357(1) | 1.00 | 2.87 |
| Gd$_{0.9}$Ir$_4$Sb$_{10.2}$Sn$_{1.8}$ | 9.28884(5) | 0.1530(7) | 0.3403(7) | 0.70 | 1.85 |
| Yb$_{0.9}$Ir$_4$Sb$_{10.2}$Sn$_{1.8}$ | 9.27823(6) | 0.1533(8) | 0.3405(9) | 0.57 | 1.89 |

Figure1S. The Rietveld refinements of the laboratory powder XRD data for filled skutterudites (and comparison to binary skutterudites). All patterns were collected with Cu-kα at 300 K.

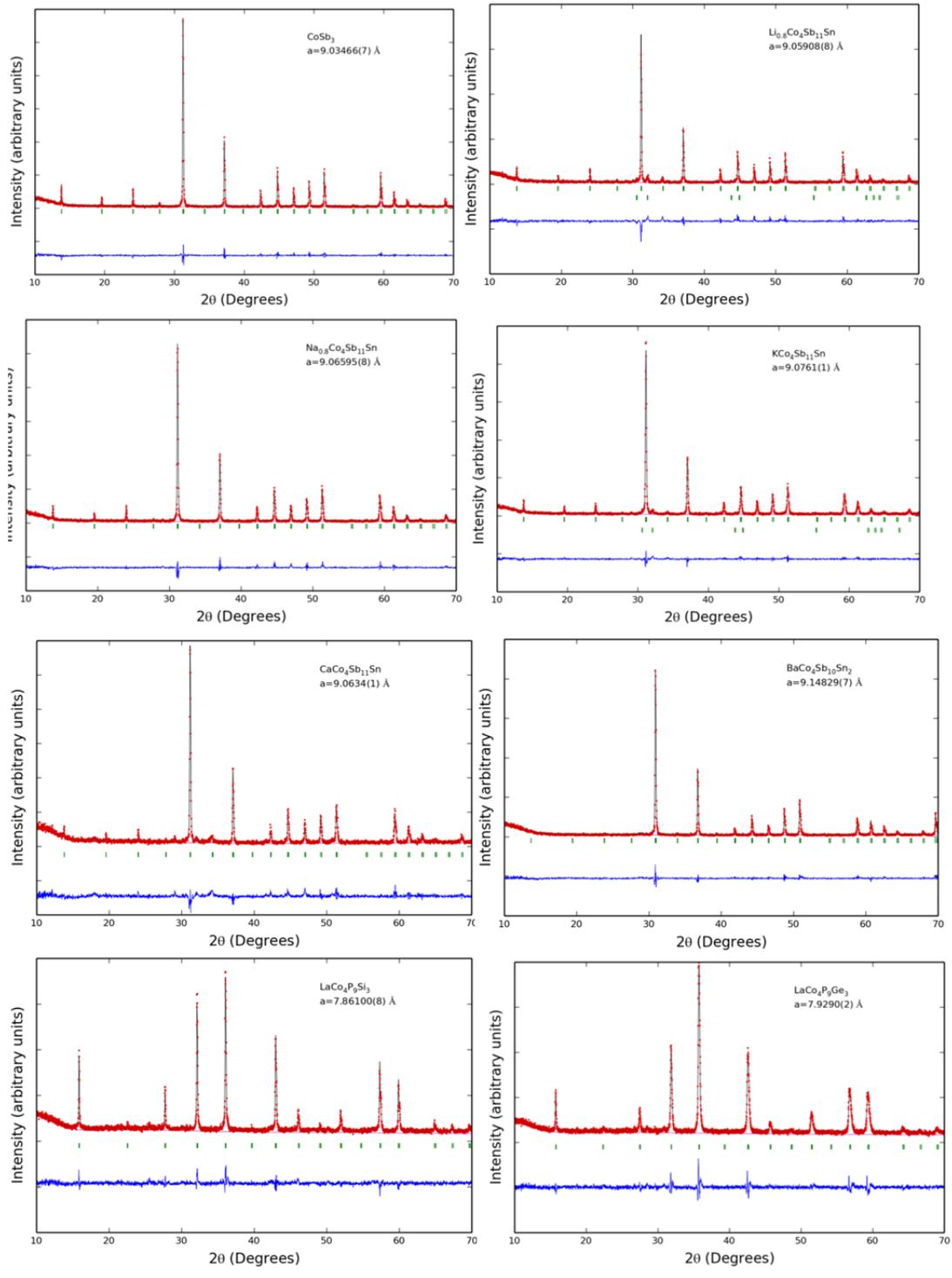

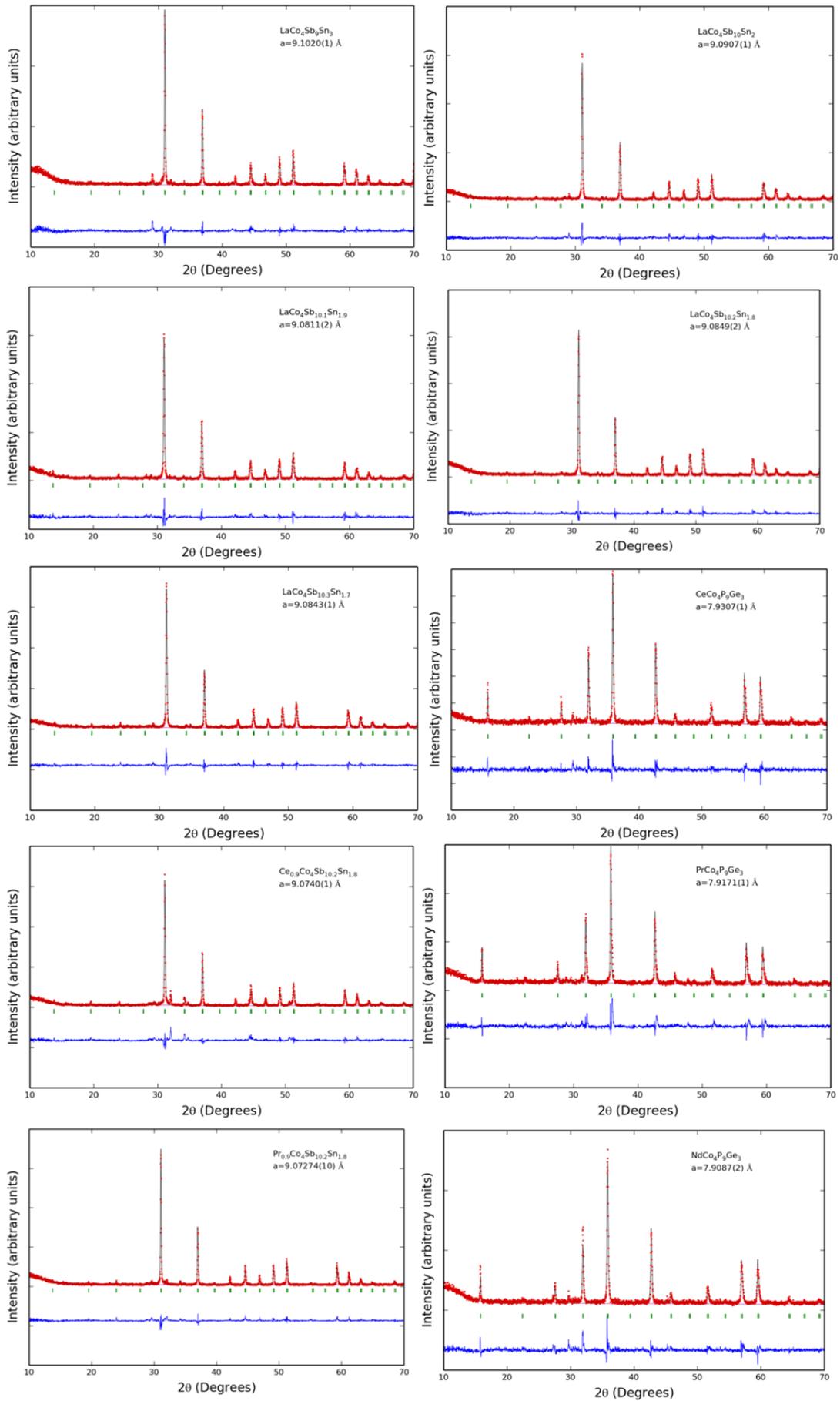

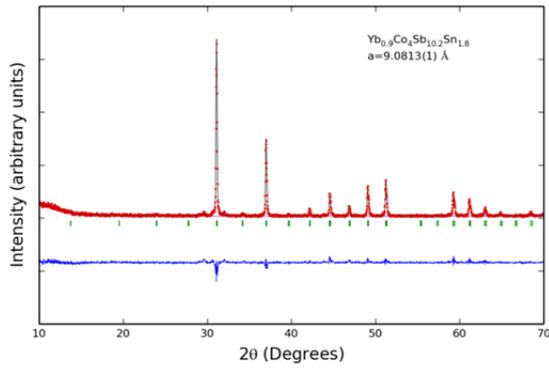
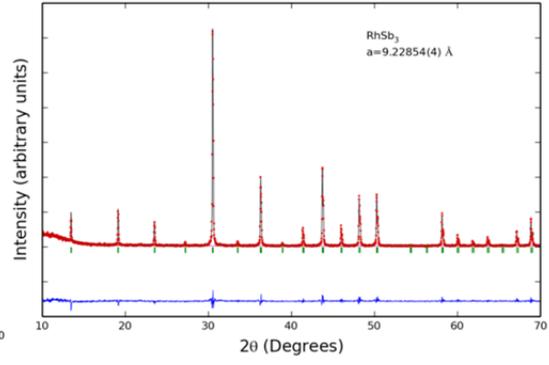
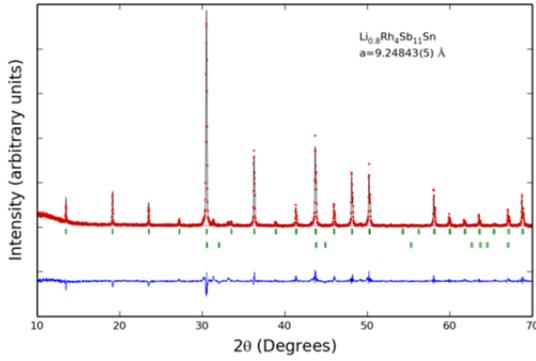
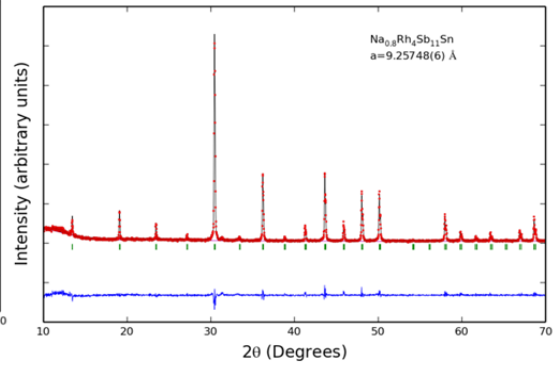
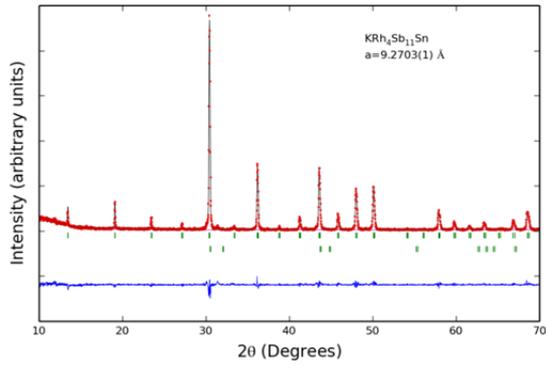
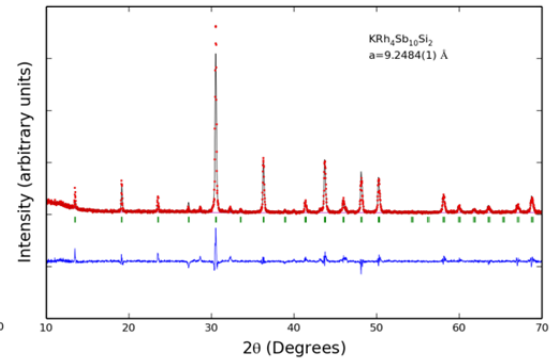
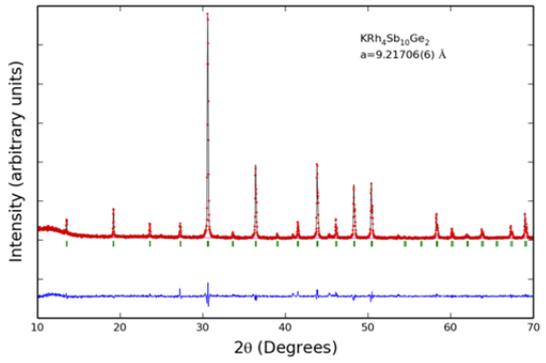
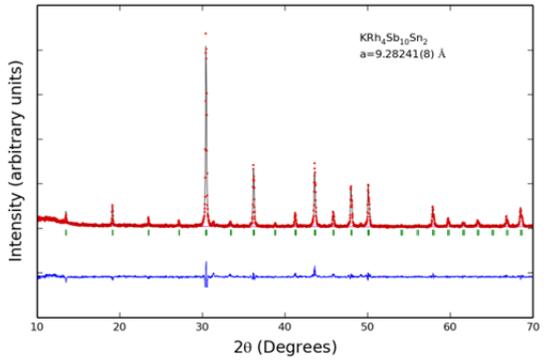
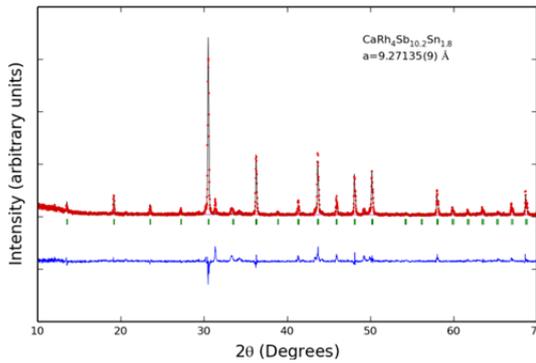
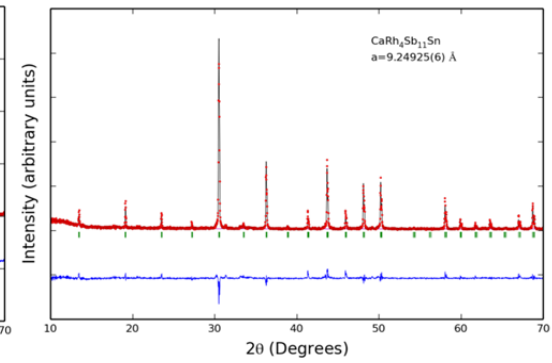

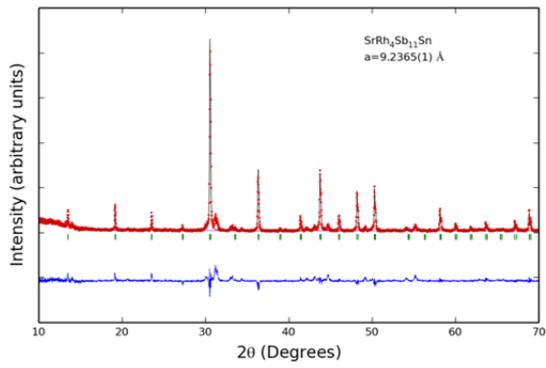
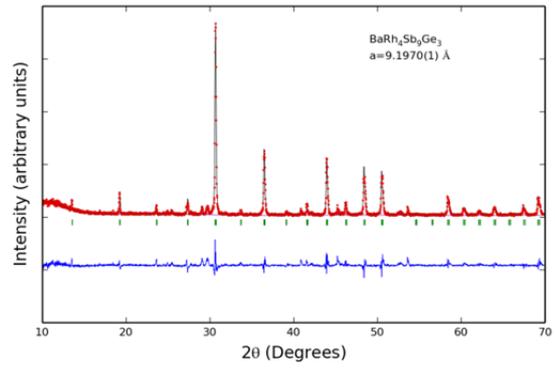
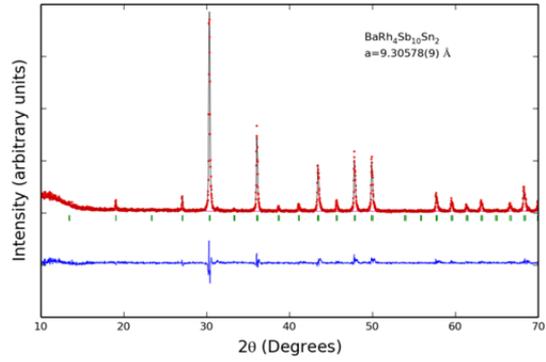
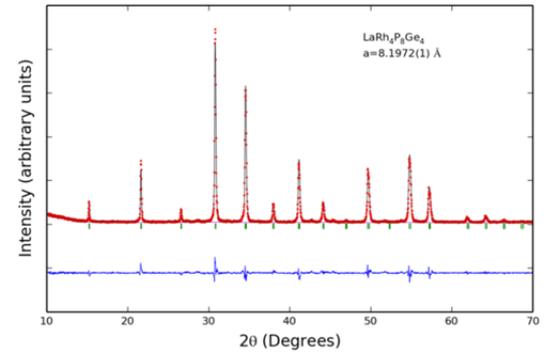
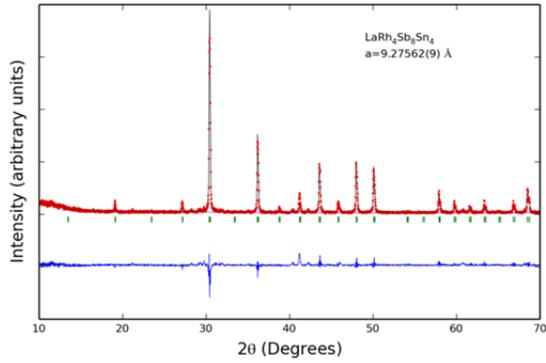
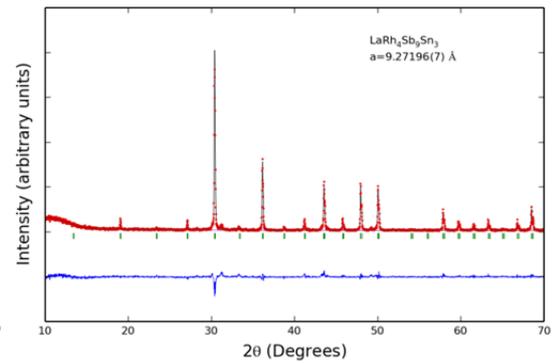
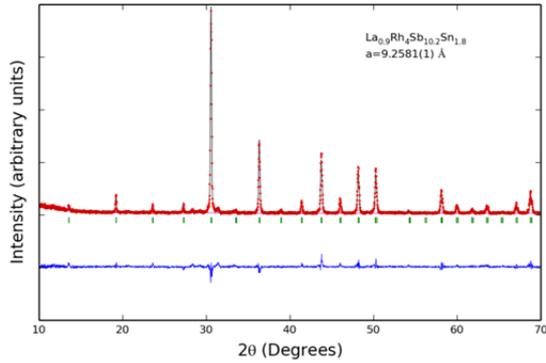
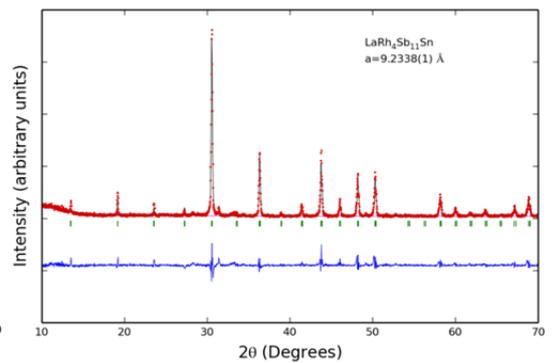
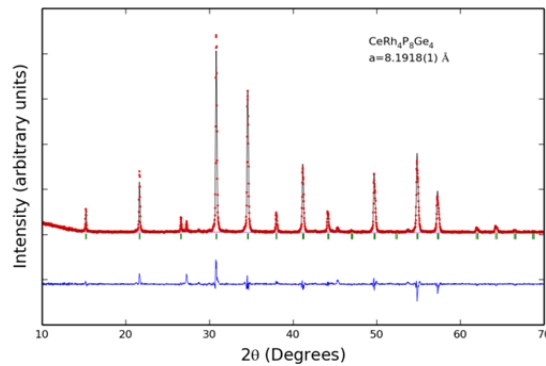
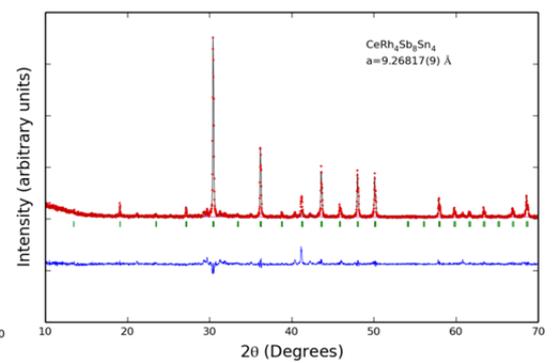

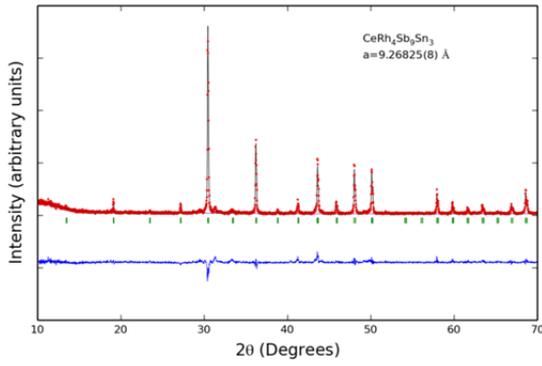
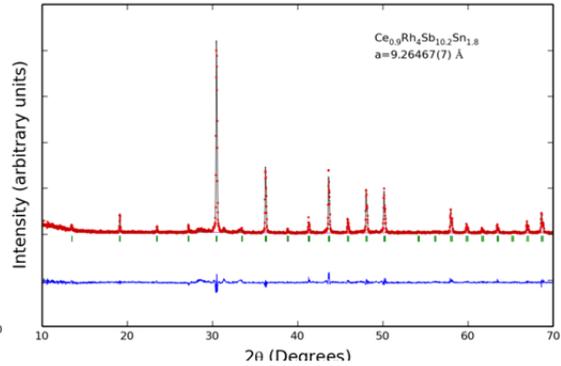
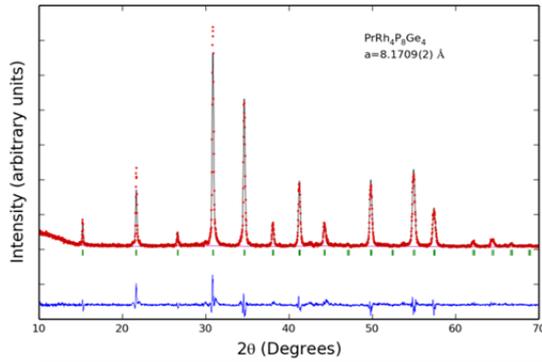
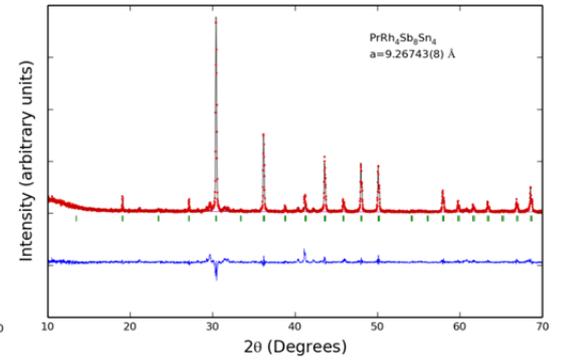
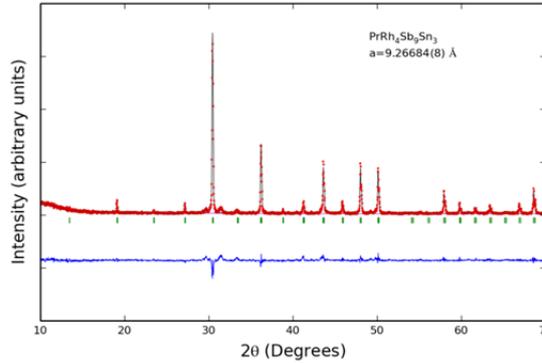
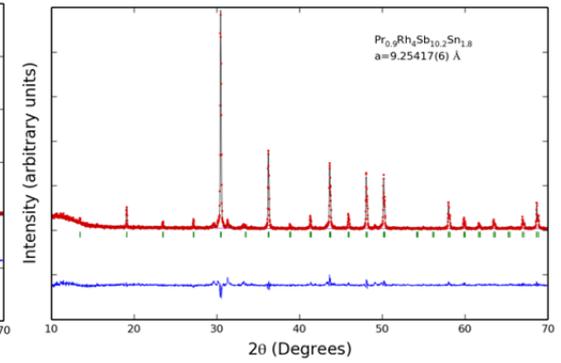
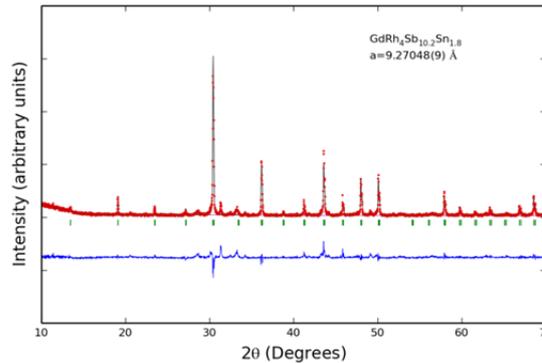
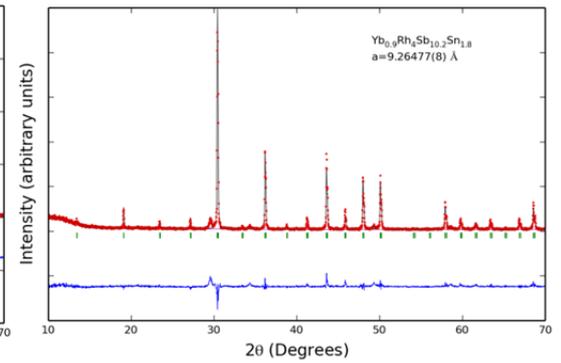
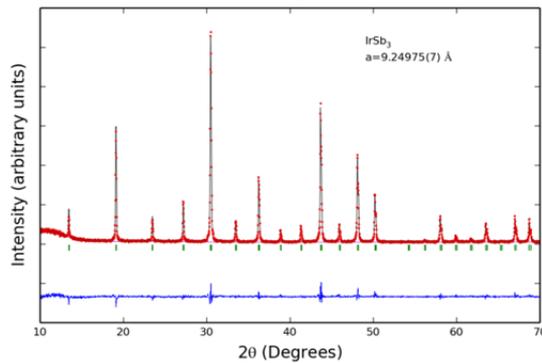
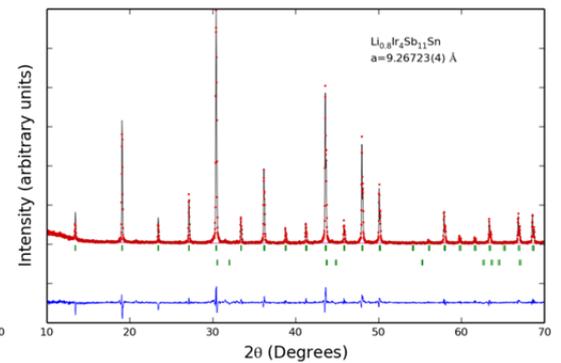

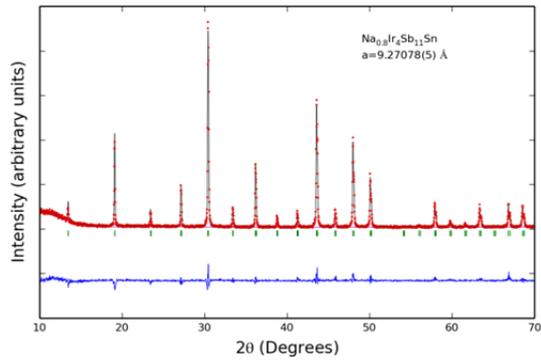
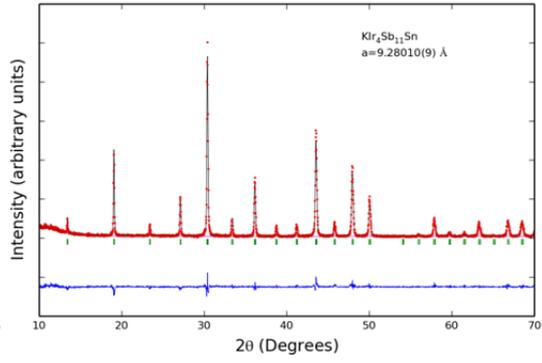
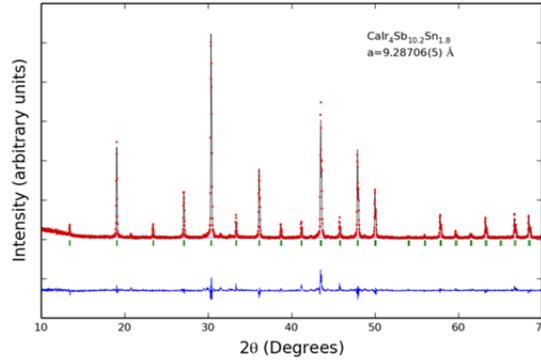
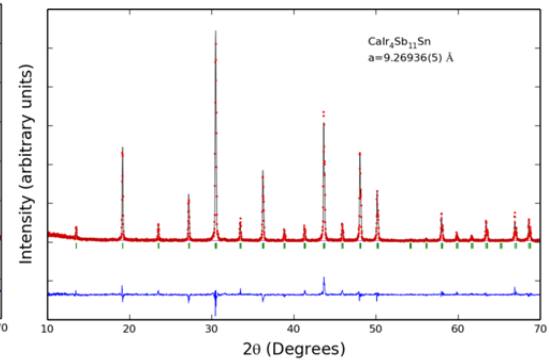
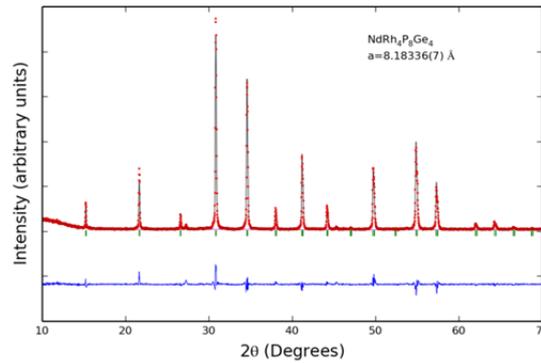
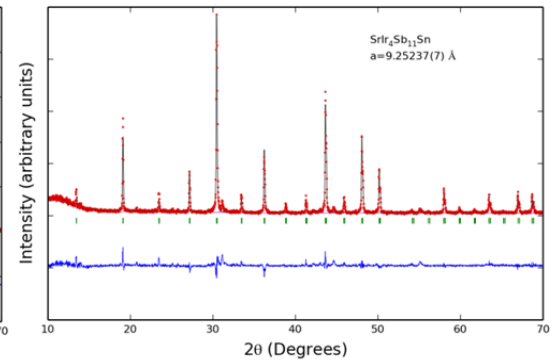
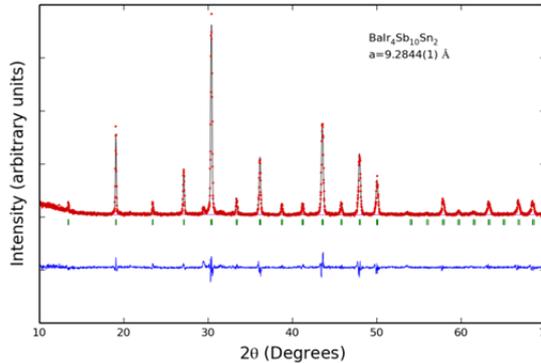
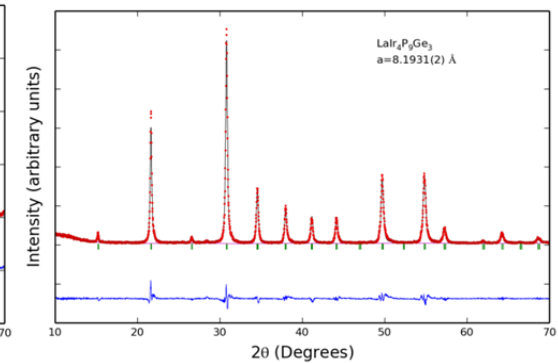
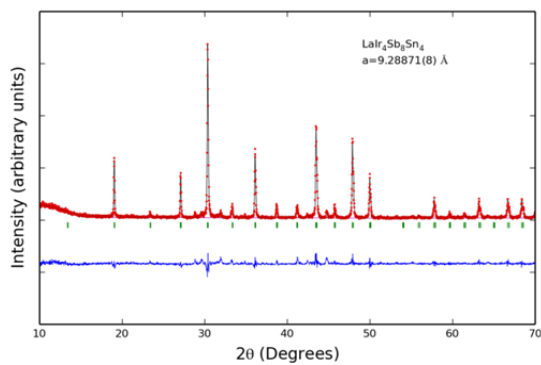
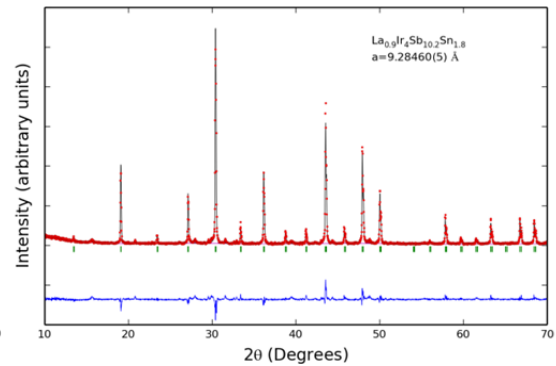

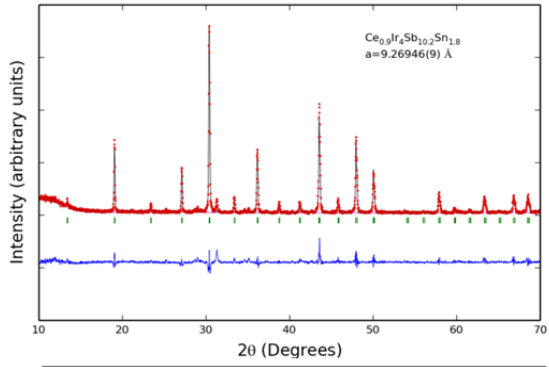
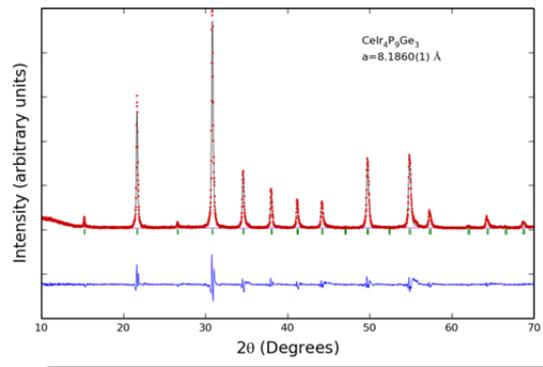
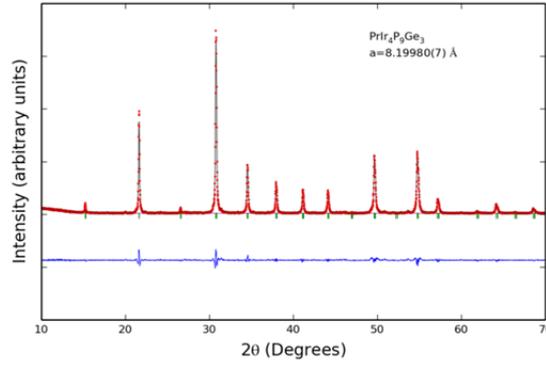
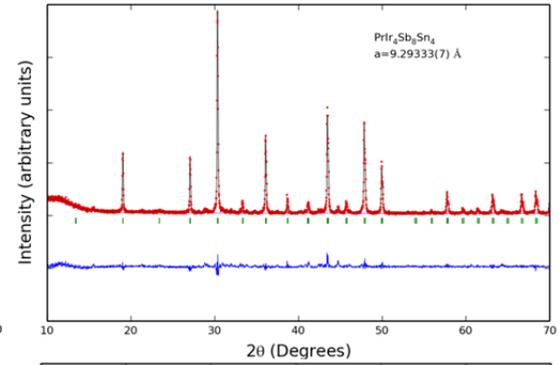
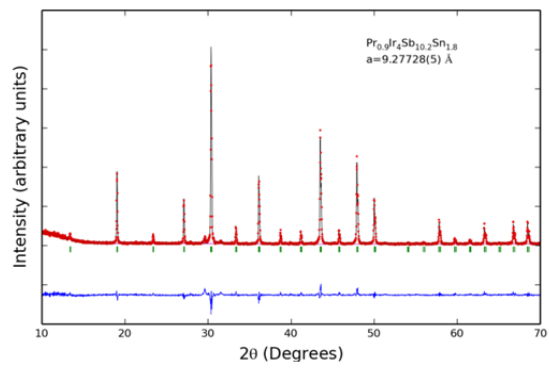
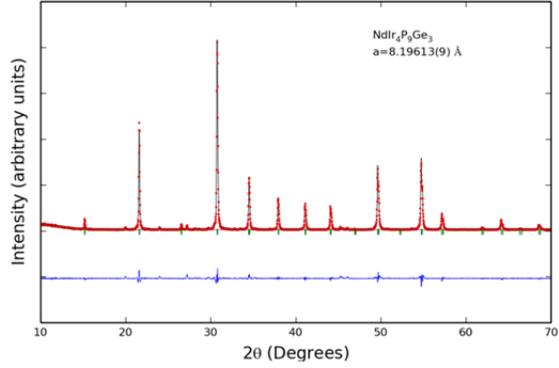
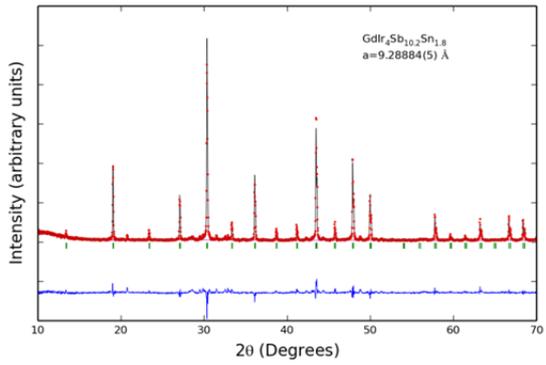
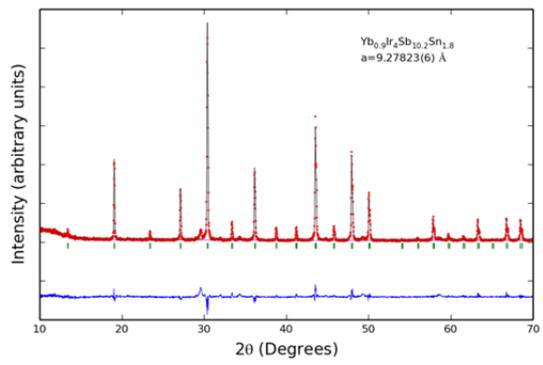